\documentclass[12pt]{article}
\usepackage{amsmath,amssymb}
\usepackage{graphicx}
\usepackage{color}
\usepackage{subfigure}

\begin{document}
\baselineskip=16pt
\begin{titlepage}
\begin{flushright}
{\small SU-HET-03-2013}\\
{\small EPHOU-13-007}\\
{\small IPMU13-0171}
\end{flushright}
\vspace*{1.2cm}

\begin{center}

{\Large\bf Next to new minimal standard model} 
\lineskip .75em
\vskip 1.5cm

\normalsize
{\large Naoyuki Haba}$^{1,2}$,
{\large Kunio Kaneta}$^{2,3,4}$ and 
{\large Ryo Takahashi}$^2$

\vspace{1cm}
$^1${\it Graduate School of Science and Engineering, Shimane University, 

Matsue, Shimane 690-8504, Japan}\\
$^2${\it Department of Physics, 
Faculty of Science, Hokkaido University, 

Sapporo, Hokkaido 060-0810, Japan}\\
$^3${\it Kavli IPMU (WPI), 
The University of Tokyo,\\ Kashiwa, Chiba 277-8568, Japan}\\
$^4${\it Department of Physics, Graduate School of Science, Osaka 
University, 

Toyonaka, Osaka 560-0043, Japan}\\

\vspace*{10mm}

{\bf Abstract}\\[5mm]
{\parbox{13cm}{\hspace{5mm}
We suggest a minimal extension of the standard model, 
 which can explain current experimental data of  
 the dark matter, 
 small neutrino masses and baryon asymmetry of the universe, 
 inflation,
 and dark energy, 
 and achieve gauge coupling unification. 
The gauge coupling unification  
 can explain the charge quantization, 
 and be realized by introducing 
 six new fields.  
We investigate 
 the vacuum stability,  
 coupling perturbativity,  
 and correct dark
 matter abundance  
 in this model by use of 
 current experimental data.}}

\end{center}

\end{titlepage}

\section{Introduction}
The standard model (SM) in particle physics has achieved great success in the 
last few decades. In particular, a recent discovery of the Higgs particle with 
the mass of 126 GeV at the CERN Large Hadron Collider (LHC) 
experiment\cite{Chatrchyan:2013lba} filled the last piece of the SM. So far the
 results from the LHC experiment are almost consistent with the SM, and no 
signatures of new physics such as the supersymmetry (SUSY) or extra-dimension(s)
  are discovered. However, there are some unsolved problems in the SM, for 
example, there is no candidate of dark matter (DM) in the SM, which are expected
 to be solved by the new physics beyond the SM. 
 
The SUSY is an excellent candidate for the physics beyond the SM 
 since it 
 solves the gauge hierarchy problem and realizes
 the gauge coupling 
 unification (GCU) as well as contains 
 the DM candidate. 
But, 
 the recent 
 discovery of the Higgs with the $126$
 GeV mass and no signature of the 
 SUSY may disfavor the SUSY at low energy. 
Actually, the magnitude 
 of the fine-tuning in the gauge hierarchy problem
 is much less than that of the
 cosmological constant problem. 
So it should be meaningful to 
 reconsider the minimum extension 
 of the SM by forgetting about 
 the gauge hierarchy problem.
A model suggested 
 in Ref.\cite{Davoudiasl:2004be} 
 was a minimal extension of the
 SM,
\footnote{See also~\cite{Chpoi:2013wga} and references therein for related works.} which is called new minimal SM (NMSM). 
In addition to the SM fields, 
 the NMSM contains 
 a gauge singlet scalar,
 two right-handed neutrinos, an 
 inflaton, and the small cosmological constant, 
 which can explain the DM, 
 small neutrino 
 masses and baryon asymmetry of the universe (BAU), 
 inflation, and dark energy (DE), respectively. 
Although a favored parameter space in the NMSM 
 for the vacuum stability, triviality bounds, and the 
 correct DM abundance was shown in Ref.\cite{Davoudiasl:2004be}, 
 experimental data was old. For example, 
 the allowed region for the scalar singlet 
 DM is also updated~\cite{Cline:2013gha} by utilizing the results of the 
 LHC searches for invisible Higgs decays, the thermal relic density of the DM, 
 and DM searches via indirect and direct detections, recently.
The parameter search must be investigated again 
 with the current experimental data. 
This is one motivation of this paper. 

It is worth noting that 
 the GCU can not be achieved in the NMSM. 
The charge quantization is one of the 
 biggest problems in the SM, 
 which should be solved in a grand unified theory (GUT). The GCU can be a 
sufficient condition of the GUT, and the great merit of the SUSY SM is just the 
realization of the GCU. Thus, here we suggest next to new minimal SM (NNMSM) in 
order to achieve the GCU by extending the NMSM. Our model includes six new 
fields, two adjoint fermions and four vector-like $SU(2)_L$ doublet fermions, in
 addition to the particle contents of the NMSM. We also revisit the stability 
and triviality bounds with the $126$ GeV Higgs mass, the recent updated 
limits on the DM particle, and the latest experimental value of the top pole 
mass as 173.5 GeV. The vacuum stability and triviality bounds are quit sensitive 
to the Higgs and top masses. 
We will point out that there are parameter 
 regions in which the stability and triviality bounds,
 the correct abundance of DM, and the 
 Higgs and top masses can be realized at the same time. 

\section{Next to new minimal standard model}

We suggest next to new minimal standard model (NNMSM) by extending the
 NMSM, which has  
 the gauge singlet real scalar boson $S$,
 two right-handed neutrinos $N_i$,
 the inflaton $\varphi$,
 and the small cosmological constant $\Lambda$ 
 in addition to the SM. 
Our
 model introduces six new fields such as two adjoint fermions 
 $\lambda_a~(a=2,3)$ and four vector-like $SU(2)_L$-doublet fermions,  
 $L'_i$ and $\overline{L'_i}~(i=1,2)$,
 in addition to the particle contents of the NMSM.
The quantum numbers of these particles are given in Table.~\ref{1}, where the 
quantum number of $L'_i$ and $\overline{L'_i}$ is the same as that of the SM 
lepton doublet.\footnote{Other possibilities for particle contents are studied in Ref.\cite{Haba:2013dva}} The gauge singlet scalar and two adjoint fermions have 
odd-parity under an additional $Z_2$-symmetry while other additional particles 
have even-parity. We will show the singlet scalar becomes DM as in the NMSM. 
Runnings of gauge couplings are changed from the SM due to new particles with 
the charges. The realization of the GCU is one of important results of this work
 as we will show later. 

We consider the NNMSM as a renormalizable theory, and thus, the relevant 
Lagrangian of the NNMSM is given by
\begin{eqnarray}
 {\cal L}_{\rm NNMSM}   
  &=& {\cal L}_{\rm SM} + {\cal L}_S +{\cal L}_N
      +{\cal L}_\varphi+{\cal L}_\Lambda+{\cal L}', \\
 {\cal L}_{\rm SM} 
  &\supset& -\lambda\left(|H|^2-\frac{v^2}{2}\right)^2, \label{SM} \\ 
 {\cal L}_S 
  &=& -\bar{m}_S^2S^2-\frac{k}{2}|H|^2S^2 - \frac{\lambda_S}{4!}S^4
      +(\text{kinetic term}), \label{S} \\
{\cal L}_N
 &=& -\left(\frac{M_i}{2}\overline{N_i^c}N_i
            +h_\nu^{i\alpha}\overline{N_i}L_\alpha\tilde{H}+c.c.\right)
     +(\text{kinetic term}), \label{neu} \\
 {\cal L_\varphi}
  &=& -B\varphi^4\left[\mbox{ln}\left(\frac{\varphi^2}{\sigma^2}\right)
                       -\frac{1}{2}\right]-\frac{B\sigma^4}{2}
      -\mu_1\varphi|H|^2
      -\mu_2\varphi S^2-\kappa_H\varphi^2|H|^2
      -\kappa_S\varphi^2S^2 \nonumber \\ 
  & & -(y_N^{ij}\varphi \overline{N_i}N_j
        +y_3\varphi\overline{\lambda_3}\lambda_3
        +y_2\varphi\overline{\lambda_2}\lambda_2
        +y_{\varphi L'}^{ij}\varphi\overline{L'_i}L'_j+c.c.)
        +(\text{kinetic term}), \label{inf} \\
 {\cal L}_\Lambda
  &=& (2.3\times10^{-3}\mbox{ eV})^4, \label{cc} \\
 {\cal L}' &=& 
  \left[(y_L^{i\alpha}L'_i\tilde{H}+y_{\bar{L}}^{i\alpha}\overline{L'_i}^\dagger H^\dagger)E_\alpha
  +M_3\overline{\lambda_3}\lambda_3+M_2\overline{\lambda_2}\lambda_2
  +M_{L'_i}\overline{L'_i}L'_i+h.c.\right] \nonumber \\
  & &+(\text{kinetic terms}), \label{Yukawa} 
\end{eqnarray}
with $\alpha=e,\mu,\tau$ and $\tilde{H}=i\sigma_2H^\ast$ where ${\cal L}_{SM}$ 
is the Lagrangian of the SM, which includes the Higgs potential.
$v$ is the vacuum expectation value (VEV) of the Higgs as $v=246$ GeV. 
${\cal L}_{S,N,\varphi,\Lambda}$ are Lagrangians
 for the dark matter, 
 right-handed neutrinos, inflaton, and the cosmological constant,
 respectively.  
${\cal L}_{\rm SM}+{\cal L}_{S,N,\Lambda}$ are the same as those of 
the NMSM.\footnote{For the present cosmic acceleration, we simply assume that 
the origin of DE is the tiny cosmological constant, which is given in 
$\mathcal{L}_\Lambda$ of Eq.(\ref{cc}), so that the NNMSM predicts the equation 
of state parameter as $\omega=-1$, like the NMSM. We will not focus on the DE in
 this work anymore.} ${\cal L}'$ is new Lagrangian in the NNMSM, where $E$ is 
right-handed charged lepton in the SM. Mass matrix, $M_{L}$, is assumed to be 
diagonal, for simplicity. 
\begin{table}[tbp]
\begin{center}
\begin{tabular}{c|ccccccc} \hline \hline
           & $\lambda_3$ & $\lambda_2$ & $L'_i$  & $\overline{L'_i}$ & $S$  & $N_i$ & $\varphi$ \\ \hline
 $SU(3)_C$ & 8           & 1           & 1      & 1                & 1    & 1         & 1     \\
 $SU(2)_L$ & 1           & 3           & 2      & 2                & 1    & 1         & 1     \\
 $U(1)_Y$  & 0           & 0           & $-1/2$ & $1/2$            & 0    & 0         & 0     \\ \hline
 $Z_2$     & $-$         & $-$         & $+$    & $+$              & $-$  & $+$       & $+$   \\ \hline \hline
\end{tabular}
\end{center}
\caption{Quantum numbers of additional particles ($i=1,2$).}
\label{1}
\end{table}

There are several mass scales of new particles, i.e., masses of DM, right-handed
 neutrinos, adjoint fermions, and inflaton. For the minimal setup, we introduce 
two mass scales in addition to the EW (TeV) scale. One is the mass of the new 
particles, $M_{\rm NP}$, and all new fermions have the mass scale as 
$M_3\simeq M_2\simeq M_{L'_i}\simeq M_{\rm NP}$. The other is the scalar DM with
 the mass  
 \begin{eqnarray}
  m_S=\sqrt{\bar{m}_S^2+kv^2/8},
 \end{eqnarray}
which is constrained by experiments and a realization of the correct abundance 
of the DM. Actually, there are other options for the setup of building the 
model, which  will be shown later. 

\subsection{Gauge coupling unification}

At first, we investigate the runnings
 of the gauge couplings in the NNMSM. 
Since we
 introduce two adjoint fermions $\lambda_3$ and $\lambda_2$, 
 and four vector-like 
 $SU(2)_L$-doublet fermions, $L'_i$ and $\bar{L}_i$ $(i=1,2)$,
 listed in Tab.~\ref{1}, 
 the beta functions of the RGEs for the gauge couplings become 
 \begin{eqnarray}
  2\pi\frac{d\alpha_j^{-1}}{dt} &=& b_j^{\rm SM}+b_j', \label{g}
 \end{eqnarray}
where 
 $(b_1^{\rm SM},b_2^{\rm SM},b_3^{\rm SM})=(41/10,-19/6,-7)$ for  
 the SM,
 and $(b_1',b_2',b_3')=(4/5,8/3,2)$ for new 
 contributions in the NNMSM. 
$t\equiv\ln(\mu/1\mbox{ GeV})$ and $\mu$ is the
 renormalization scale, and 
 $\alpha_j\equiv g_j^2/(4\pi)$ $(j=1,2,3)$ with
 $g_1\equiv\sqrt{5/3}g'$.  
Since all masses of new particles 
 are around the same scale, 
 $\Lambda_{\rm EW}<M_{\rm NP}\simeq M_3\simeq M_2\simeq M_{L'_i}$,
 where $\Lambda_{\rm EW}$ is the EW scale,  
 we should utilize the RGEs of 
 Eq.(\ref{g}) at high energy scale ($M_{NP}\leq\mu$) 
 while the right-handed side of
 Eq.(\ref{g}) must be $b_j^{\rm SM}$ 
 at low energy scale ($\Lambda_{\rm EW}\leq\mu<M_{NP}$). 

\begin{figure}
\begin{center}
\includegraphics[scale=1,bb=0 0 270 180]{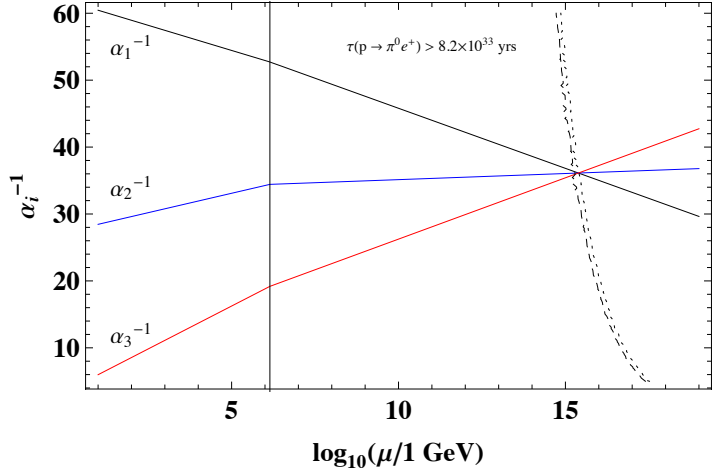}
\end{center}
\caption{The runnings of the gauge couplings in the NNMSM.
The horizontal axis is the 
renormalization scale and
 the vertical axis is the values of $\alpha_i^{-1}$. 
The runnings of $\alpha_1^{-1}$, $\alpha_2^{-2}$, and $\alpha_3^{-1}$ are 
described by black, blue, and red solid curves, respectively. 
We take $M_{\rm NP} = 1.40\times 10^3$ TeV, and the coupling unification is
 realized at $\mu=\Lambda_{\rm GUT}\simeq 2.45\times 10^{15}$ GeV with
 $
 \alpha_{\rm GCU}^{-1}\simeq
 36.1$.
Dotted (dashed) contour shows
 the experimental limit of $p\to\pi^0 e^+$ as
 $\tau(p\to\pi^0e^+)=8.2\times10^{33}$, 
 by use of 
 $\alpha_H=-0.0146~ (-0.0078~{\rm GeV}^3)$. 
}
\label{fig2}
\end{figure}

According to the numerical analyses, 
 taking a free parameter $M_{NP}$ as 
 $
 1.40\times10^3$ TeV
 can realize
 the GCU with a good precision at 
 1-loop level as shown in Fig.~\ref{fig2}.\footnote{
In this analysis, we take the following values as~\cite{Beringer:1900zz},
$
  \sin^2\theta_W(M_Z)=0.231,\alpha_{\rm em}^{-1}(M_Z)=128,
  \alpha_s(M_Z)=0.118,
$
for the parameters in the EW theory, where $\theta_W$ is the Weinberg angle, 
$\alpha_{em}$ is the fine structure constant, and $\alpha_s$ is the strong 
coupling, respectively. 
} 
We show the threshold of new particles with $1.40\times10^3$ TeV mass by a black
 solid line. The NNMSM suggests the GCU at 
 \begin{eqnarray}
  \Lambda_{\rm GCU}\simeq2.45\times10^{15}\mbox{ GeV} \label{lambda_GCU}
 \end{eqnarray} 
with the unified coupling as 
 \begin{eqnarray} 
  \alpha_{\rm GCU}^{-1}\simeq36.1. \label{alpha_GCU}
 \end{eqnarray} 
Suppose the minimal $SU(5)$ GUT at $\Lambda_{\rm GCU}$, the protons
decay of $p\to\pi^0e^+$ occurs by exchanging heavy gauge bosons of the
GUT gauge group, and here we estimate a limit from the proton life time.
A constraint from the proton decay experiments is 
$\tau(p\rightarrow\pi^0e^+)>8.2\times10^{33}$ years~\cite{Beringer:1900zz}, and 
 the partial decay width of proton for $p\rightarrow\pi^0e^+$ is given by 
 \begin{eqnarray}
  \Gamma(p\rightarrow\pi^0e^+)
   =\alpha_H^2\frac{m_p}{64\pi f_\pi^2}(1+D+F)^2
    \left(\frac{4\pi\alpha_{\rm GCU}}{\Lambda_{\rm GCU}}A_R\right)^2(1+(1+|V_{ud}|^2)^2),
 \end{eqnarray} 
where $\alpha_H^2$ is the hadronic matrix element, $m_p$ is the proton mass, 
$f_\pi$ is the pion decay constant, $D$ and $F$ are the chiral Lagrangian 
parameters, $A_R$ is the renormalization factor, and $V_{ud}$ is a element of 
the CKM matrix (e.g., see~\cite{Ibe:2009gt,Hisano:2000dg}). 
In our analysis, we take these parameters as 
$m_p=0.94~{\rm GeV},~f_\pi=0.13~{\rm GeV}, A_R\simeq0.93,~D=0.80$ and
$F=0.47$. A theoretical uncertainty on the proton life time comes mainly from the
 hadronic matrix element as $\alpha_H=-0.0112\pm 0.0034~{\rm 
GeV}^3$~\cite{Aoki:2008ku}. When $\alpha_H=-0.0146~{\rm GeV^3}$, which is the 
lowest value, the proton life time is evaluated as $\tau\simeq 
5.7\times10^{33}$ years. On the other hand, when $\alpha_H=-0.0078~{\rm GeV^3}$,
 which is the largest value, the proton life time is $2.0\times10^{34}$ years. 
As for the center value, $\alpha_H=-0.0112~{\rm GeV^3}$, the proton life time is
 $\tau\simeq9.7\times10^{33}$ years. Thus, the NNMSM can be consistent with the 
proton decay experiment, although the conservative limit can not. In 
Fig.\ref{fig2}, dotted (dashed) contour shows the experimental limit, 
$\tau(p\to\pi^0e^+)=8.2\times10^{33}$, by use of Eqs.(\ref{lambda_GCU}) and 
(\ref{alpha_GCU}) with $\alpha_H=-0.0146~(-0.0078)~{\rm GeV^3}$. Since the 
future Hyper-Kamiokande experiment is expected to exceed the life time 
$\mathcal{O}(10^{35})$ years~\cite{Abe:2011ts}, which corresponds to 
$\Lambda_{\rm GCU}\simeq4.39_{-0.72}^{+0.62}\times10^{15}$ GeV for 
$\alpha_H=-0.0112\pm 0.0034~{\rm GeV}^3$, the proton decay is observed if the 
NNMSM is true. 

Here let us examine other numbers of 
 $L'_i$ and 
 $\overline{L'_i}$. 
When only one pair of $L'_i$ and 
 $\overline{L'_i}$ is introduced,
 the GCU is never realized even with taking $M_{NP}$ as any 
 other scales. 
If we introduce more pairs of $L'_i$ and $\overline{L'_i}$ than 
 two,
 a heavier mass scale of new particles 
 ($M_{NP}\gg10^3$ TeV) can also 
 realize the GCU. 
For examples, 
 three (four) pairs of
 $L'_i$ and $\overline{L'_i}$ with 
 $M_{NP}\simeq4.26\times10^8$ GeV
 ($7.67\times10^9$ GeV) 
 realize the GCU at 
 $\Lambda_{\rm GCU} = 6.87\times10^{14}$ GeV 
 ($3.62 \times10^{14}$ GeV).  
However, these cases cannot
 satisfy the constraint on the proton stability, and  
 introduction of more pairs of $L'_i$ and $\overline{L'_i}$ 
 leads to smaller $\Lambda_{\rm GCU}$.
Thus, we conclude two pairs of $L'_i$ and $\overline{L'_i}$ 
 is consistent with the phenomenology.

We have taken the initial setup that there are 
 two adjoint fermions and all new fermions 
 have the same scale masses. 
Under this condition, 
 the above field content is the
 minimal as the 
 the NNMSM.  
However, there are other initial setups 
 for building the NNMSM.  
One is introducing 
 different mass scales for 
 the new fermions. 
For examples, 
 the GCU can be achieved by 
 different mass scales between
 $M_3$ and $M_2$ \cite{Ibe:2009gt}. In this case, we do not need $L'_i$ and 
$\overline{L'_i}$. It has less degrees of freedom of the fields but contains 
deferent mass scales. Another is introducing several generations of adjoint 
fermions with $M_{NP}\sim10^8$ GeV, where the GCU can also be realized. This 
initial setup can induce tiny neutrino mass without two right-handed neutrinos 
like the NMSM and NNMSM, through the type-III seesaw mechanism. Taking these 
initial setups is alternative way of constructing ``another'' NNMSM, which will 
be investigated in a separate publication~\cite{HKT}.

\subsection{Abundance and stability of new fermions}

Next, we discuss an abundance and stability of
 new fermions,
 $\lambda_3$, $\lambda_2$, and $L'_i$, $\overline{L'_i}$. 
$\lambda_{3}$ and $\lambda_{2}$ are expected to be long lived 
 since they cannot decay into the SM sector 
 due to the $Z_2$-symmetry. 
A stable colored particle is
 severely constrained by experiments with heavy 
 hydrogen isotopes,  
 since it bounds in nuclei and appears 
 as anomalously heavy isotopes
 (e.g., see~\cite{Giudice:2004tc}). The number of the stable colored 
 particles per nucleon should be smaller
 than $10^{-28}~(10^{-20})$ for its mass 
 up to 1 (10) TeV~\cite{Smith:1982qu,Hemmick:1989ns}. 
But the calculation of
 the relic abundance of the stable colored
 particle is uncertain because of
 the dependence on the mechanism of hadronization and nuclear 
 binding\cite{Baer:1998pg}. 

In this paper,
 we apply a simple scenario in order to avoid the problem of the 
 presence of the stable colored particle. 
It is to consider few production 
 scenario for the stable particle, i.e.,
 the stable particles were rarely 
 produced in the thermal history of the universe and 
 clear  
 the constraints of the colored particles. 
In fact, 
 a particle with mass of $M$ is very rarely produced 
 thermally 
 if the reheating temperature after the 
 inflation is lower than $M/(35\sim40)$.\footnote{We 
 thank S. Matsumoto for pointing out it 
 in a private discussion.} 
Therefore, we consider a relatively low reheating temperature as 
 \begin{eqnarray}
  T_{RH}\lesssim\frac{M_{NP}}{40}=25\mbox{ TeV},
 \end{eqnarray}
since $M_{NP}=10^3$ TeV. $\lambda_2$ is also rarely
 produced in the thermal history of 
 the universe. 
Therefore, the presence of two new adjoint fermions in the NNMSM 
 for the GCU is not problematic. The vector-like fermions $L_{i}$ and 
 $\overline{L'_i}$ are also rarely produced  
 (if they are produced, they  
 decay into the SM particles through the Yukawa interactions 
 in Eq.(\ref{Yukawa}) before the Big Bang nucleosynthesis (BBN)). 
Therefore Yukawa couplings of ${L'_i}$ 
 in Eq.(\ref{Yukawa}) are not constrained. 

If we introduces an additional gauge singlet fermion $N'$  
 with odd parity, 
 the decay of $\lambda_3$ can be induced through 
 dimension-6 operator,
 $\frac{\lambda_6}{\Lambda^2}\overline{Q}Q\lambda_3N'$. 
However, in order for $\lambda_3$ to decay before
 the BBN, $\Lambda<\mathcal{O}(10^{13})$ GeV is required. 
We consider the 
 NNMSM as the renormalizable theory, and 
 we do not want to introduce this new scale 
 which could induces various higher dimensional operators. 
Thus, we do not introduce the above operator in 
 the NNMSM. 

\subsection{Inflation}

Next, we discuss the inflation, and 
 the relevant Lagrangian is given by 
 $\mathcal{L}_\varphi$ in Eq.(\ref{inf}). 
The 
WMAP~\cite{Bennett:2012zja,Hinshaw:2012aka} and the Planck~\cite{Ade:2013zuv} 
measurements of the cosmic microwave background (CMB) constrain the cosmological
 parameters related with the inflation in the early universe. In particular, the
 first results based on the Planck measurement with a WMAP polarization 
low-multipole likelihood at $\ell\leq23$ 
(WP)~\cite{Bennett:2012zja,Hinshaw:2012aka} and high-resolution (highL) CMB data
 gives
 \begin{eqnarray}
  n_s &=& 0.959\pm0.007~(68\%;~\mbox{Planck$+$WP$+$highL}), \\
  r_{0.002} &<& \left\{ 
                 \begin{array}{ll}
                  0.11 & (95\%;~\mbox{no running},~\mbox{Planck$+$WP$+$highL}) \\
                  0.26 & (95\%;~\mbox{including running},~\mbox{Planck$+$WP$+$highL})
                 \end{array}
                \right., \\
  dn_s/d\mbox{ln}k &=& -0.015\pm0.017~(95\%;~\mbox{Planck$+$WP$+$highL}),
 \end{eqnarray}
for the scalar spectrum power-law index, the ratio of tensor primordial power to
 curvature power, the running of the spectral index, respectively, in the 
context of the $\Lambda$CDM model. 
Regarding 
 $r_{0.002}$, the constraints are given for both no running and including running
 cases of the spectral indices.

In the NMSM, the relevant Lagrangian for the inflation is given by 
 \begin{eqnarray}
 {\cal L_\varphi}=-\frac{m^2}{2}\varphi^2-\frac{\mu}{3!}\varphi^3
                  -\frac{\kappa}{4!}\varphi^4.
 \end{eqnarray}  
If the inflaton starts with a trans-Planckian amplitude, the model corresponds 
to the chaotic inflation model~\cite{Linde:1983gd}. The benchmark point 
discussed in~\cite{Davoudiasl:2004be} was $m\simeq1.8\times10^{13}$ GeV, 
$\mu\lesssim10^6$ GeV, and $\kappa\lesssim10^{-14}$. Since the terms 
$\frac{\mu}{3!}\varphi^3$ and $\frac{\kappa}{4!}\varphi^4$ are dominated by the 
quadratic term of $\frac{m^2}{2}\varphi^2$ at this point, this inflation model 
is similar to the simplest inflation model with a quadratic potential. This type
 of the inflationary model can be on the absolute edge of the constraint from 
the Planck (95\%) when the $e$-folds is $N\simeq60$~\cite{Ade:2013zuv}. The 
values of coupling of inflaton with the Higgs, DM, and right-handed neutrinos, 
are appropriately chosen by the reheating temperature for the thermal 
leptogenesis~\cite{Fukugita:1986hr} and keeping the flatness of the inflaton 
potential. However, we must require the relatively low reheating temperature as 
$T_{RH}\lesssim25$ TeV for the few production scenario of additional fermions. 
Such a low reheating temperature leads to smaller $e$-folds as $N<60$ in the 
chaotic inflation, which lies outside the joint 95\% CL for 
Planck$+$WP$+$highL data. In order to realize the $e$-folds as $60\lesssim N$ 
in this chaotic inflation model, $4\times10^{16}$ GeV$\lesssim T_{RH}$ should be
 taken.

Therefore, we adopt a different inflation model for the NNMSM, which is given by
 $\mathcal{L}_\varphi$ in Eq.(\ref{inf}). 
The inflaton potential is the 
 Coleman-Weinberg (CW) type\cite{Coleman:1973jx,Knox:1992iy},
 which is generated
 by radiative corrections. 
In this potential Eq.(\ref{inf}),  the VEV of $\varphi$ 
becomes $\sigma$. When we take $(\phi,\sigma,B)\simeq(6.60\times10^{19}\mbox{ 
GeV},9.57\times10^{19}$ GeV$,10^{-15})$, the model can lead to $n_s=0.96$, 
$r=0.1$, $dn_s/d\mbox{ln}k\simeq8.19\times10^{-4}$, and 
$(\delta\rho/\rho)\sim\mathcal{O}(10^{-5})$, which are consistent with the 
cosmological data. The values of couplings of inflaton with the Higgs, DM, 
right-handed neutrinos, and new fermions are also constrained because there is 
an upper bound on the reheating temperature after the inflation as 
$T_{RH}\lesssim25$ TeV. This upper bound leads to 
$\mu_{1,2}\lesssim9.23\times10^3$ GeV and $(y_N^{ij},y_3,y_2,y_{\varphi 
L'}^{ij})\lesssim2.41\times10^{-10}$. Since $\kappa_{H,S}$ should be almost 
vanishing at the low energy for the realizations of the EW symmetry breaking and
 the DM mass, we take the values of $\kappa_{H,S}$ as very tiny at the epoch of 
inflation. The smallness of $\kappa_{H,S}$ does not also spoil the stability and
 triviality bounds, which will be discussed in the next section. As for the 
lower bound of the reheating temperature, it depends on the baryogenesis 
mechanism. When the baryogenesis works through the sphaleron process, the 
reheating temperature must be at least higher than $\mathcal{O}(10^2)$ GeV. 

There are a large number of inflation models even in the context of 
single-field inflationary models (e.g., see~\cite{Tsujikawa:2013ila} for the 
Planck constraints on single-field inflation),
 so it is interesting to 
 investigate whether
 other inflation models can be embedded into the NNMSM. 
Where we should consider or construct an inflation model satisfying 
 non-trivial constraints in the NNMSM in addition to the cosmological data. 
These are that the inflationary model must (i) realize low reheating temperature
 for the tiny abundance of the adjoint fermions (upper bound), and (ii) take 
coupling constant(s) to the scalar sector of the NNMSM as small enough not to 
spoil the stability and triviality conditions, EW symmetry breaking, and the DM 
mass. As mentioned above, the upper bound on the reheating temperature in the 
inflation model depends on the mass scale of the new particles for the GCU. If 
one can realize the GCU with different particle contents and the corresponding
 mass scale, there might be other possible inflation and suitable baryogenesis 
models.

\section{Stability, triviality, 
 dark matter, neutrino, and baryogenesis}

In this section, 
 we investigate parameter region where 
 not only stability and triviality bounds but also 
 correct abundance of the DM 
 are achieved. 
Realizations of the suitable tiny active neutrino mass and 
 baryogenesis are also discussed. 

\subsection{Stability, triviality, and dark matter}

The ingredients of Higgs and DM sector
 in the NNMSM is the same as
 the NMSM\cite{Davoudiasl:2004be}, 
 which are given by $\mathcal{L}_{\rm SM}$ and 
 $\mathcal{L}_S$ in Eqs.(\ref{SM}) and (\ref{S}). The singlet scalar $S$ becomes
 the DM. In Ref.\cite{Davoudiasl:2004be}, 
 the Higgs 
 boson mass was predicted to be in the range of
 $130$ GeV$\lesssim m_h\lesssim180$ GeV for values
 of $\lambda_S(M_Z)=0,1,1.2$ with  
top Yukawa coupling $y(M_Z)=1$
(corresponding to the top $\overline {\rm MS}$ mass
 $m_t(M_Z)\simeq174$ GeV). However, 
 the stability and triviality bounds  
 are very sensitive to
 the top mass, and then, 
 it is important to reanalyze
 the stability and triviality bounds with
 the $126$ GeV Higgs mass and 
 the latest experimental value of the top
 pole mass\cite{Beringer:1900zz,CDF:2013jga},
\begin{equation}
M_t=173.5\pm 1.4\  {\rm GeV}.
\label{M_t}
\end{equation}
We should also use 
 the present limits for the singlet DM model. 

The RGEs for three quartic couplings of the scalars are given 
 by~\cite{Davoudiasl:2004be},
 \begin{eqnarray}
  (4\pi)^2\frac{d\lambda}{dt} 
   &=& 24\lambda^2+12\lambda y^2-6y^4-3\lambda(g'{}^2+3g^2)
   +\frac{3}{8}\left[2g^4+(g'{}^2+g^2)^2\right]+\frac{k^2}{2}, \label{lam} \\
  (4\pi)^2\frac{dk}{dt} 
   &=& k\left[4k+12\lambda+\lambda_S+6y^2-\frac{3}{2}(g'{}^2+3g^2)\right], 
   \label{k} \\
  (4\pi)^2\frac{d\lambda_S}{dt} &=& 3\lambda_S^2+12k^2. \label{h}
 \end{eqnarray}
We comment on Eq.(\ref{k}) that right-hand side of the equation is
proportional to $k$ itself.
Thus, if we take a small value of $k(M_Z)$, evolution of $k$ tends to be
slow and remained in a small value, and the running of $\lambda$ closes
to that of SM. In our analysis, boundary conditions of the Higgs self-coupling 
and top Yukawa coupling are given by
\begin{eqnarray}
  \lambda(M_Z)=\frac{m_h^2}{2v^2}=0.131,~~~
  y(M_t)=\frac{\sqrt{2}m_t(M_t)}{v} \label{BC}
 \end{eqnarray}
for the RGEs, where the vacuum expectation value of the Higgs field is
$v=246~{\rm GeV}$.

\begin{figure}
\subfigure[]{
\includegraphics[clip,width=0.464\columnwidth]{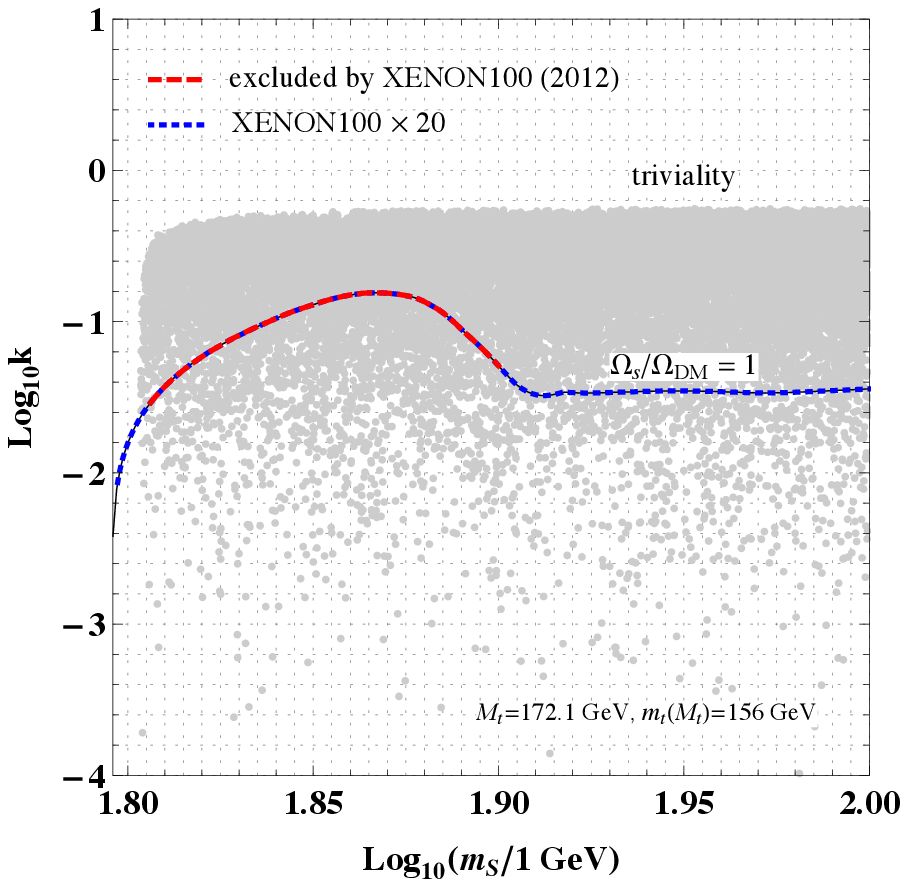}
}
\subfigure[]{
\includegraphics[clip,width=0.45\columnwidth]{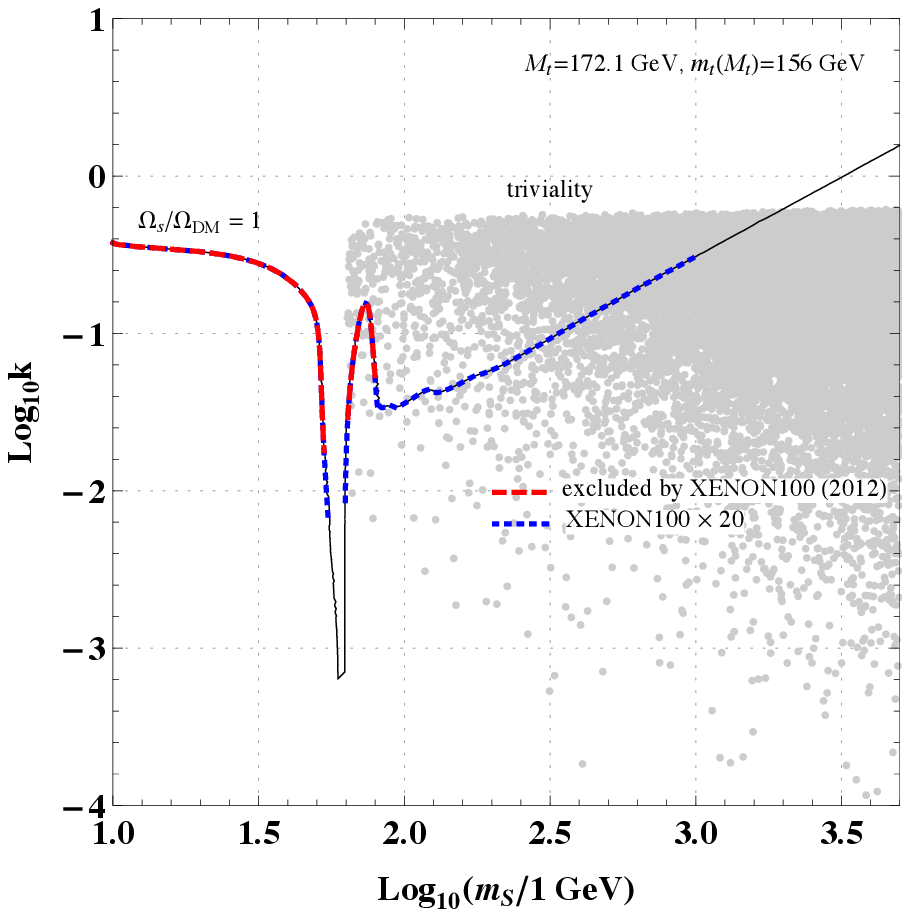}
}
\caption{
A contour of fixed relic density $\Omega_S/\Omega_{\rm DM}=1$
 and a region, wihch is described by gray plots, satisfying the stability and
 triviality bounds with $M_t=172.1$ GeV ($m_t(M_t)=156$
 GeV). 
The upper boundary of the gray region is determined by
 the triviality condition, where ``triviality'' in the both figures represents the corresponding condition.
The (red) dashed and (blue) 
 dotted lines are experimental limits from XENON100 (2012) and 20 times 
sensitivity of XENON100, respectively.
 (a) The mass region is 
 $63~{\rm GeV}\leq m_S \leq 100~{\rm GeV}$
 ($1.8 \leq \log(m_S/1~{\rm GeV})\leq 2.0$).
 (b) The mass region is 
 $10~{\rm GeV}\leq m_S \leq 5000~{\rm GeV}$
 ($1.0 \leq \log(m_S/1~{\rm GeV})\leq 3.7$). 
}
\label{omega_Mt1721}
\end{figure}

Let us solve the RGEs, Eqs.(\ref{lam})$\sim$(\ref{h}),
 and obtain the stable 
 solutions, i.e.,
 the 
 scalar
 quartic couplings are 
 within the range of
 $0<(\lambda,k,\lambda_S)<4\pi$ up to the Planck scale 
 $M_{\rm pl}=10^{18}$ GeV. 
Figure \ref{omega_Mt1721} shows the case of $M_t=172.1~{\rm GeV}$
 (corresponding to $m_t(M_t)=156~{\rm GeV}$), 
 which is the smallest value of the top pole mass 
 in Eq.(\ref{M_t}). 
The solutions of the RGEs are described by gray plots in 
 Fig.~\ref{omega_Mt1721}, where
 the horizontal and vertical axes are
$\log_{10}(m_S/1\mbox{ GeV})$ and $\log_{10}k$ at the $M_Z$ scale.
The smaller top pole mass becomes, 
 the smaller top Yukawa contribution in
 Eq.(\ref{lam}) becomes.  
Thus the stability bound 
 tends to be relaxed by comparing 
 larger top mass cases, 
 and actually 
 this case does not suffer from 
 stability condition. 
We also show the contour satisfying $\Omega_S/\Omega_{\rm DM}=1$ with 
 $\Omega_{\rm DM}=0.115$,  
 where $\Omega_S$ and $\Omega_{\rm DM}$ are density 
 parameter of the singlet DM and observed value of the 
 parameter~\cite{Bennett:2012zja}, respectively. 
The contour is calculated by
{\tt micrOMEGAs}~\cite{Belanger:2013oya}. 
Since there is no other DM candidate except for the $S$ to
compensate $\Omega_S/\Omega_{\rm DM}<1$, which is above the contour, we
focus only on the contour.
The relic density depends on $k$ and $m_S$ but not $\lambda_S$, 
meanwhile $\lambda_S$ affects the stability and triviality bounds. In the 
figure, $\lambda_S(M_Z)$ is randomly varied from 0 to $4\pi$, where 
$\lambda_S$-dependence of the stability and triviality bounds is not stringent, 
and most of $\lambda_S(M_Z) \in [0,1]$ as the boundary condition can satisfy the
  bounds. A direct DM search experiment, XENON100 (2012), gives an exclusion 
limit~\cite{Cline:2013gha}, which is described by the (red) dashed line in 
Fig.~\ref{omega_Mt1721}. There are two regions, $R_{1,2}$, which satisfy both 
the correct DM abundance and the triviality bound simultaneously, 
\begin{eqnarray}
 R^{(M_t=172.1)}_1
&\hspace{-3mm}=\hspace{-3mm}&\left\{
\begin{array}[tb]{l l}
63.5~{\rm GeV}\lesssim m_S \lesssim 64.0~{\rm GeV} &
 (1.803\lesssim\log_{10}(m_S/1~{\rm GeV})\lesssim 1.806)\\
2.40\times10^{-2}\lesssim k(M_Z) \lesssim 2.63\times10^{-2} &
 (-1.64\lesssim\log_{10}k(M_Z)\lesssim -1.58)
\end{array}
\right.,\nonumber \\ && \\
R^{(M_t=172.1)}_2
&\hspace{-3mm}=\hspace{-3mm}&\left\{
\begin{array}[tb]{l l}
81.3~{\rm GeV}\lesssim m_S \lesssim 2040~{\rm GeV} &
 (1.91\lesssim\log_{10}(m_S/1~{\rm GeV})\lesssim 3.31)\\
3.16\times10^{-2}\lesssim k(M_Z) \lesssim 6.31\times10^{-1} &
 (-1.50\lesssim\log_{10}k(M_Z)\lesssim -0.20)
\end{array}
\right..
\end{eqnarray}
The future XENON100 experiment with 20 times sensitivity, 
 which is described by the (blue) dotted lines
 in Fig.~\ref{omega_Mt1721}, will be able 
 to rule out the lighter $m_S$ region, $R_1$,
 completely. 
On the other hand, the
 heavier $m_S$ region, $R_2$,
 can be currently allowed by all experiments searching for 
 DM. 
It is seen that the future XENON100$\times$20 can check up to 
 $m_S\lesssim$ 1000 GeV 
 ($\log_{10}(m_S/1\mbox{ GeV})\lesssim3$). 
The future XENON1T experiment and combined data from indirect detections 
of Fermi$+$CTA$+$Planck at $1\sigma$ CL may be able to reach up to $m_S\simeq5$ 
TeV~\cite{Cline:2013gha}.

Next, let us show 
 the heaviest top pole mass, $M_t=174.9~{\rm GeV}$,  
 in Eq.(\ref{M_t}), where 
 the allowed regions become narrow as 
\begin{eqnarray}
 R^{(M_t=174.9)}_1 &\hspace{-3mm}=\hspace{-3mm}& R^{(M_t=172.1)}_1,\\
R^{(M_t=174.9)}_2
&\hspace{-3mm}=\hspace{-3mm}&\left\{
\begin{array}[tb]{l l}
1862~{\rm GeV}\lesssim m_S \lesssim 2040~{\rm GeV} &
 (3.27\lesssim\log_{10}(m_S/1~{\rm GeV})\lesssim 3.31)\\
5.74\times10^{-1}\lesssim k(M_Z) \lesssim 6.31\times10^{-1} &
 (-0.24\lesssim\log_{10}k(M_Z)\lesssim -0.20)
\end{array}
\right.. \nonumber \\
\end{eqnarray}
This is because the larger top Yukawa coupling
 gives stringent bound on the vacuum stability.\footnote{
The 
 NMSM~\cite{Davoudiasl:2004be} 
 predicted the larger Higgs mass region
 as $130$ GeV$\lesssim m_h\lesssim180$ GeV. 
It is because 
 the top mass was taken as 
 $m_t(M_Z)=174$ GeV, and  
 such the large top Yukawa coupling 
 induces vacuum instability. 
} 
On the other hand, the small $m_S$ region, $R_1$, 
 does not change from the case of $M_t = 172.1$ GeV. 
The reason is as follows. 
In the RGE analyses, 
 $k$ in the R.H.S. of Eq.(\ref{lam}) is effective above 
 the energy scale of $m_S$. 
Then, the triviality bound of $\lambda$ becomes severe as 
 the $m_S$ becomes small, and 
 the left-edge of gray dots shows this bound.   
This does not depend on the top Yukawa coupling, 
 so that the region $R_1$ is independent of 
 the top pole mass.  

\begin{figure}
\subfigure[]{
\includegraphics[clip,width=0.464\columnwidth]{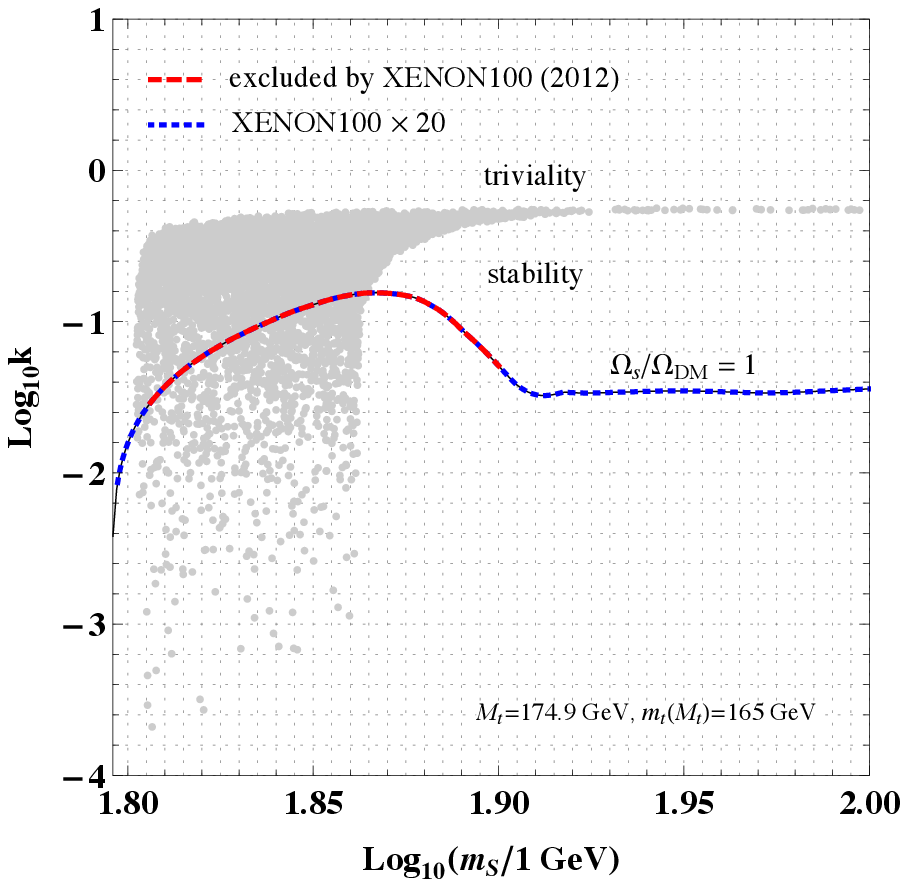}
}
\subfigure[]{
\includegraphics[clip,width=0.45\columnwidth]{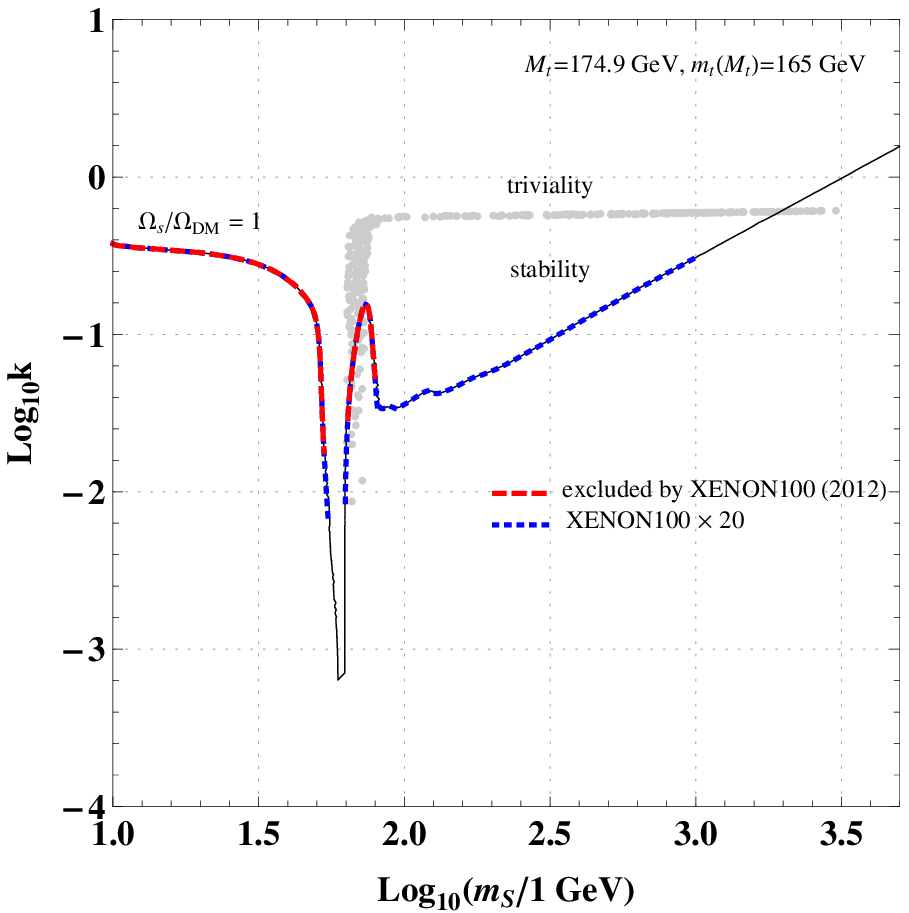}
}
\caption{
The same plots as Fig.\ref{omega_Mt1721} with 
 $M_t=174.9$ ($m_t(M_t)=165$ GeV). 
The lower and upper boundaries of the gray region are determined by
 the stability and triviality conditions, respectively, where ``stability'' and
 ``triviality'' in the both figures represent the corresponding conditions.
(a) The mass region is 
 $63~{\rm GeV}\leq m_S \leq 100~{\rm GeV}$
 ($1.8 \leq \log(m_S/1~{\rm GeV})\leq 2.0$). 
(b) The mass region is 
 $10~{\rm GeV}\leq m_S \leq 5000~{\rm GeV}$
 ($1.0 \leq \log(m_S/1~{\rm GeV})\leq 3.7$). 
}
\label{omega_Mt1749}
\end{figure}

\begin{figure}
\subfigure[]{
\includegraphics[clip,width=0.464\columnwidth]{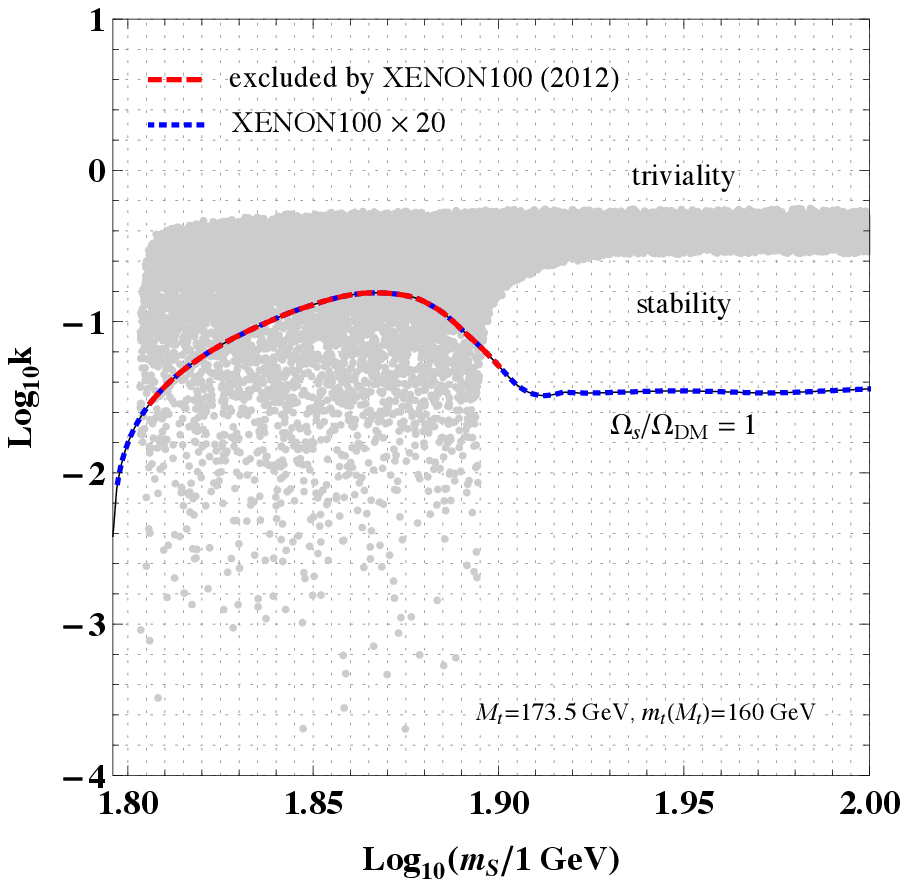}
}
\subfigure[]{
\includegraphics[clip,width=0.45\columnwidth]{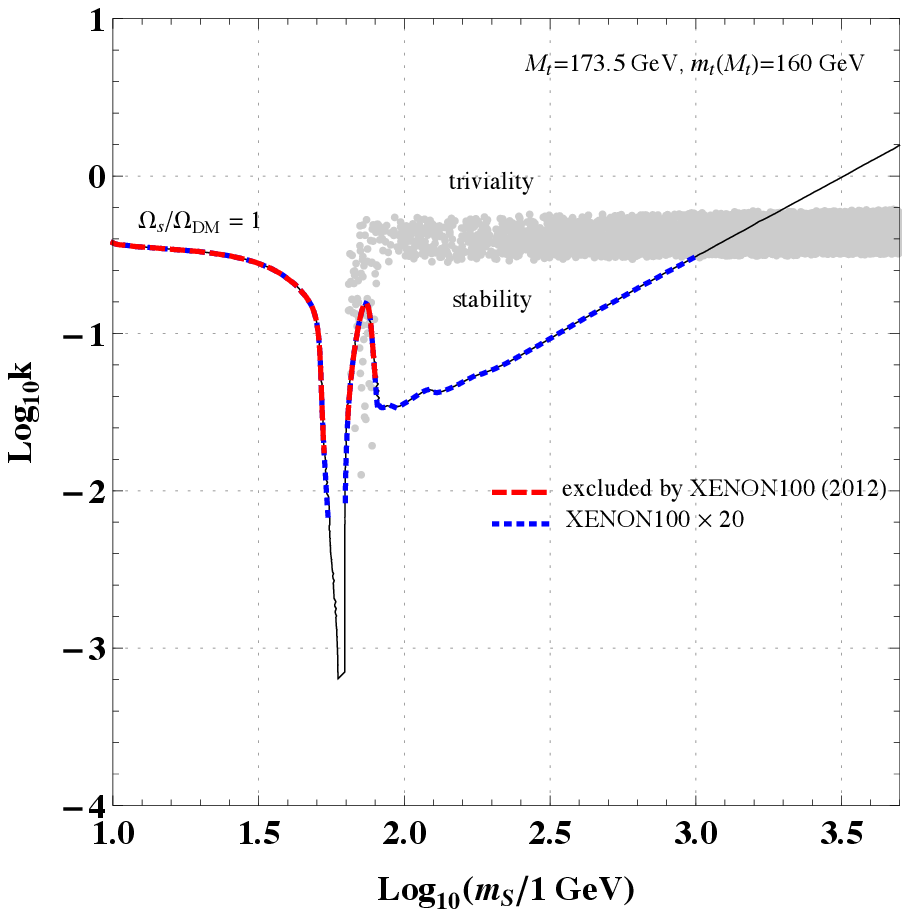}
}
\caption{The same plots as Fig.\ref{omega_Mt1721} with 
  $M_t=173.5$ ($m_t(M_t)=160$ GeV). 
(a) The mass region is
 $63~{\rm GeV}\leq m_S \leq 100~{\rm GeV}$
  ($1.8 \leq \log(m_S/1~{\rm GeV})\leq 2.0$). 
(b) The mass region is 
  $10~{\rm GeV}\leq m_S \leq 5000~{\rm GeV}$
 ($1.0 \leq \log(m_S/1~{\rm GeV})\leq 3.7$).}
\label{omega_Mt1735}
\end{figure}

Finally, 
 let us show the case of 
 the center value of the top pole mass, $M_t=173.5~{\rm GeV}$, in 
Fig.\ref{omega_Mt1735}. In the figure, we can find the regions satisfying the
 correct DM abundance and the stability and triviality bounds as, 
\begin{eqnarray}
 R^{(M_t=173.5)}_1 &\hspace{-3mm}=\hspace{-3mm}& R^{(M_t=172.1)}_1,\\
 R^{(M_t=173.5)}_2
&\hspace{-3mm}=\hspace{-3mm}&\left\{
\begin{array}[tb]{l l}
955~{\rm GeV}\lesssim m_S \lesssim 2040~{\rm GeV} &
 (2.98\lesssim\log_{10}(m_S/1~{\rm GeV})\lesssim 3.31)\\
2.94\times10^{-1}\lesssim k(M_Z) \lesssim 6.31\times10^{-1} &
 (-0.53\lesssim\log_{10}k(M_Z)\lesssim -0.20)
\end{array}
\right.. \nonumber \\
\end{eqnarray}
We can show that $R_1$ region is the same as other top pole mass cases. 
As for the region $R_2$, it is the middle of above two figures, and 
 we notice again that the top Yukawa dependence is 
 quit large. 

Here let us show a typical example of the
 RGE running of 
 scalar quartic couplings, Eqs.(\ref{lam})-(\ref{h}), 
 with $M_t=173.5$. 
\begin{figure}
\begin{center}
\includegraphics[scale=1]{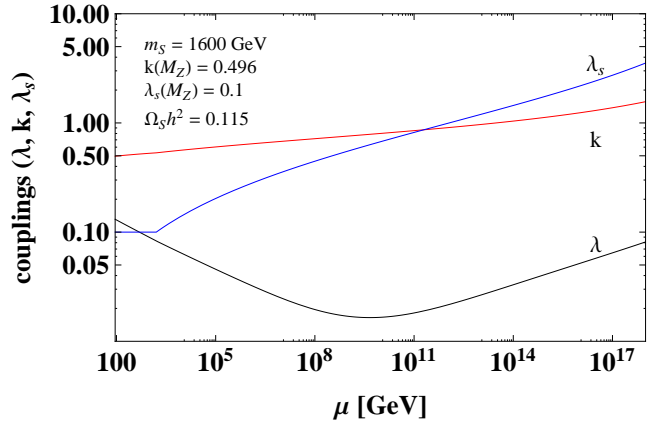}
\end{center}
\caption{An example of the stable solutions of the RGEs for scalar quartic 
couplings.}
\label{fig1}
\end{figure}
In Fig.~\ref{fig1}, 
 the horizontal axis is the renormalization scale and the vertical
 axis is the value of the scalar quartic couplings. The black, red, and blue 
solid curves indicate the runnings of $\lambda$, $k$, and 
$\lambda_S$, respectively, and 
 we take values of the couplings as 
 \begin{eqnarray}
  k(M_Z)=0.496,~~~\lambda_S(M_Z)=0.1,
 \end{eqnarray}
 with $m_S=1600$ GeV, which realize the 
 correct relic density and stability and triviality to the Planck scale. 

\subsection{Neutrinos and baryogenesis}

The neutrino sector is shown in Eq.(\ref{neu}),
 where  
 tiny active neutrino mass 
 is obtained through the type-I seesaw 
 mechanism~\cite{seesaw}. 
Since there are
 two right-handed neutrinos, one of 
 active neutrinos is predicted to be massless $m_1=0$ ($m_3=0$) for the normal 
 (inverted) mass hierarchy. 
Reminding 
 the low reheating 
 temperature 
 in the NNMSM, 
 masses of the right-handed neutrino 
 must be lighter than 
 25 TeV. What mechanism can induce the suitable baryon asymmetry 
 in such a low reheating temperature? 
One possibility is
 the resonant leptogenesis\cite{Pilaftsis:2003gt}\footnote{
It is known that the singlet DM model can induce a 
 strong EW phase transition for the EW baryogenesis 
 in some parameter regions\cite{Kuzmin:1985mm}.  
However, in the parameter regions searched in the 
 previous subsection,
 the singlet DM model cannot 
 explain total energy density of DM, and 
 requires other candidates of the DM. 
Thus, in the NNMSM, we 
 the resonant leptogenesis 
 is preferable than the EW-baryogenesis. 
} 
 in which the 
 right-handed neutrinos can be light such as
 $1$ TeV. Thus, the reheating temperature, 
 1 TeV$\lesssim T_{RH}\lesssim25$ TeV, 
 can realize the resonant leptogenesis, 
 which means the 
 couplings of inflaton as $369\mbox{ 
GeV}\lesssim\mu_{1,2}\lesssim9.23\times10^3$ GeV and $9.63\times10^{-12}\lesssim
 y_N^{ij}\lesssim2.41\times10^{-10}$ 
 in Eq.(\ref{inf}).
For the suitable light active neutrino mass, 
 neutrino Yukawa couplings should be small. 
If one allows fine-tunings among the
 neutrino Yukawa couplings, larger neutrino Yukawa couplings can also reproduce 
 experimental values in the neutrino sector in the context of the low scale 
 seesaw mechanism.\footnote{ 
In the case, the searches of the lepton flavor violating (LFV)
 processes such as $\mu\rightarrow e\gamma$, $\mu\rightarrow3e$, and 
$\mu^--e^-$ conversion may constrain and/or check the sizes of the neutrino 
Yukawa couplings (e.g., see~\cite{Kersten:2007vk,Ibarra:2011xn,Dinh:2012bp}).}

\section{Summary}

The SM has achieved great success in the last few decades,
 however, there are some 
 unsolved problems such as explanations for DM, gauge hierarchy
 problem, tiny neutrino mass scales, baryogenesis, inflation, and the
 DE. The minimally extended SM without the SUSY, so-called NMSM, 
 could explain the above problems except for the gauge hierarchy problem and GCU
  by adding two gauge singlet real scalars and two right-handed neutrinos, small
 cosmological constant. 
In this paper,
 we suggested the NNMSM 
 for the realization of the GCU by extending the NMSM. 
We take a setup that all new fermions have 
 the same mass scale of new physics. 
Under the condition, the GCU with the proton stability 
 determines the field contents of the NNMSM, i.e.,  
 six new fields such as two adjoint fermions under $SU(3)_C$ and 
 $SU(2)_L$, and four vector-like $SU(2)_L$ doublet 
 fermions are added to the 
 particle contents of the NMSM. 
The GCU can occur at $\Lambda_{\rm GCU}\simeq2.45\times10^{15}\mbox{
 GeV}$ with 
 the mass scale of the new particles
 as $10^3$ TeV. 
We consider 
 low reheating temperature, 
 $T_{RH}\lesssim25$ TeV, 
 in order not to produce the 
 stable adjoint fermions in the early universe. 
This low reheating temperature requires 
 the following issues. 
The masses of right-handed neutrino 
 should be lighter than 25 TeV, 
 so that tiny neutrino mass 
 is realized through the Type-I seesaw
 with relatively small neutrino Yukawa couplings. 
The BAU should be achieved 
 through, for example, 
 the resonant leptogenesis.  
For the 
 inflation model, 
 it should 
 (i) realize low reheating temperature, and
 (ii) take coupling constants to the scalar sector of the NNMSM
 as small enough not to 
 spoil the stability and triviality
 conditions,
 EW symmetry breaking, and the DM mass.

We have also analyzed the stability and 
 triviality conditions by use of 
 recent experimental data of Higgs and top masses. 
We found the parameter 
 regions in which the correct
 abundance of dark matter can be also
 realized at the
 same time. 
One is the lighter $m_S$ region as
 $63.5~{\rm GeV}\lesssim m_S \lesssim 64.0~{\rm GeV}$,
 and the other is 
 heavier ones as 
 $708~{\rm GeV}\lesssim m_S \lesssim 2040~{\rm GeV}$ 
 with the center value of top pole mass. 
We have shown the top mass dependence 
 is quite large even within the 
 experimental error of top pole 
 mass. 
The future XENON100 experiment with 20 times 
 sensitivity will completely check out the 
 lighter mass region. 
On the other hand, the heavier mass region 
 will also be completely checked by the 
 future 
 direct experiments of XENON100$\times$20, XENON1T and/or combined data from 
 indirect detections of Fermi$+$CTA$+$Planck at $1\sigma$ CL. 

\subsection*{Acknowledgement}

We are grateful to Shigeki Matsumoto and Osamu Seto for useful discussions. This
 work is partially supported by Scientific Grant by Ministry of  Education and 
Science, Nos. 00293803, 20244028, 21244036, 23340070, and by the SUHARA Memorial
 Foundation. The works of K.K. and R.T. are supported by Research Fellowships of
 the Japan Society for the Promotion of Science for Young Scientists.
The work is also supported by World Premier International Research Center Initiative (WPI Initiative), MEXT, Japan.




\begin{thebibliography}{99}
\bibitem{Chatrchyan:2013lba}
  G.~Aad {\it et al.}  [ATLAS Collaboration],
  Phys.\ Lett.\ B {\bf 716} (2012) 1
  [arXiv:1207.7214 [hep-ex]];
  S.~Chatrchyan {\it et al.}  [CMS Collaboration],
  arXiv:1303.4571 [hep-ex].

\bibitem{Davoudiasl:2004be}
  H.~Davoudiasl, R.~Kitano, T.~Li and H.~Murayama,
  Phys.\ Lett.\ B {\bf 609} (2005) 117
  [hep-ph/0405097].

\bibitem{Chpoi:2013wga}
  S.~Baek, P.~Ko and W.~-I.~Park,
  JHEP {\bf 1307} (2013) 013
  [arXiv:1303.4280 [hep-ph]];
  S.~Chpoi, S.~Jung and P.~Ko,
  arXiv:1307.3948 [hep-ph].



\bibitem{Cline:2013gha}
  J.~M.~Cline, K.~Kainulainen, P.~Scott and C.~Weniger,
  arXiv:1306.4710 [hep-ph].

\bibitem{Haba:2013dva}
  N.~Haba, K.~Kaneta and R.~Takahashi,
  arXiv:1309.3254 [hep-ph].

\bibitem{Beringer:1900zz}
  J.~Beringer {\it et al.}  [Particle Data Group Collaboration],
  Phys.\ Rev.\ D {\bf 86} (2012) 010001.

\bibitem{Ibe:2009gt}
  M.~Ibe,
  JHEP {\bf 0908} (2009) 086
  [arXiv:0906.4667 [hep-ph]].

\bibitem{Hisano:2000dg}
  J.~Hisano,
  hep-ph/0004266;
  K.~S.~Babu and R.~N.~Mohapatra,
  Phys.\ Lett.\ B {\bf 715} (2012) 328
  [arXiv:1206.5701 [hep-ph]].

\bibitem{Aoki:2008ku}
  Y.~Aoki {\it et al.}  [RBC-UKQCD Collaboration],
  Phys.\ Rev.\ D {\bf 78} (2008) 054505
  [arXiv:0806.1031 [hep-lat]].

\bibitem{Abe:2011ts}
  K.~Abe, T.~Abe, H.~Aihara, Y.~Fukuda, Y.~Hayato, K.~Huang, A.~K.~Ichikawa and M.~Ikeda {\it et al.},
  arXiv:1109.3262 [hep-ex].

\bibitem{HKT}
N. Haba, K. Kaneta, and R. Takahashi, in preparation.

\bibitem{Giudice:2004tc}
  G.~F.~Giudice and A.~Romanino,
  Nucl.\ Phys.\ B {\bf 699} (2004) 65
   [Erratum-ibid.\ B {\bf 706} (2005) 65]
  [hep-ph/0406088].

\bibitem{Smith:1982qu}
  P.~F.~Smith, J.~R.~J.~Bennett, G.~J.~Homer, J.~D.~Lewin, H.~E.~Walford and W.~A.~Smith,
  Nucl.\ Phys.\ B {\bf 206} (1982) 333.

\bibitem{Hemmick:1989ns}
  T.~K.~Hemmick, D.~Elmore, T.~Gentile, P.~W.~Kubik, S.~L.~Olsen, D.~Ciampa, D.~Nitz and H.~Kagan {\it et al.},
  Phys.\ Rev.\ D {\bf 41} (1990) 2074.

\bibitem{Baer:1998pg}
  H.~Baer, K.~-m.~Cheung and J.~F.~Gunion,
  Phys.\ Rev.\ D {\bf 59} (1999) 075002
  [hep-ph/9806361].

\bibitem{Bennett:2012zja}
  C.~L.~Bennett {\it et al.}  [WMAP Collaboration],
  arXiv:1212.5225 [astro-ph.CO].

\bibitem{Hinshaw:2012aka}
  G.~Hinshaw {\it et al.}  [WMAP Collaboration],
  arXiv:1212.5226 [astro-ph.CO].
    
\bibitem{Ade:2013zuv}
  P.~A.~R.~Ade {\it et al.}  [Planck Collaboration],
  arXiv:1303.5076 [astro-ph.CO].

\bibitem{Linde:1983gd}
  A.~D.~Linde,
  Phys.\ Lett.\ B {\bf 129} (1983) 177.

\bibitem{Fukugita:1986hr}
  M.~Fukugita and T.~Yanagida,
  Phys.\ Lett.\ B {\bf 174} (1986) 45.

\bibitem{Coleman:1973jx}
  S.~R.~Coleman and E.~J.~Weinberg,
  Phys.\ Rev.\ D {\bf 7} (1973) 1888.

\bibitem{Knox:1992iy}
  Q.~Shafi and A.~Vilenkin,
  Phys.\ Rev.\ Lett.\  {\bf 52} (1984) 691;
  S.~-Y.~Pi,
  Phys.\ Rev.\ Lett.\  {\bf 52} (1984) 1725;
  L.~Knox and M.~S.~Turner,
  Phys.\ Rev.\ Lett.\  {\bf 70} (1993) 371
  [astro-ph/9209006];
  Q.~Shafi and V.~N.~Senoguz,
  Phys.\ Rev.\ D {\bf 73} (2006) 127301
  [astro-ph/0603830].

\bibitem{Tsujikawa:2013ila}
  S.~Tsujikawa, J.~Ohashi, S.~Kuroyanagi and A.~De Felice,
  arXiv:1305.3044 [astro-ph.CO].

\bibitem{CDF:2013jga}
  F.~Deliot {\it et al.}  [ATLAS and D0 Collaborations],
  arXiv:1302.0830 [hep-ex];
  CDF [Tevatron Electroweak Working Group and D0 Collaborations],
  arXiv:1305.3929 [hep-ex].

\bibitem{Belanger:2013oya}
  G.~Belanger, F.~Boudjema, A.~Pukhov and A.~Semenov,
  arXiv:1305.0237 [hep-ph].



\bibitem{seesaw}
P.~Minkowski,
Phys.~Lett. {\bf B67} (1977) 421;
T.~Yanagida, in Proceedings of the Workshop on Unified Theories
and Baryon Number in the Universe, eds.\ O.~Sawada and A.~Sugamoto
(KEK report 79-18, 1979); M.~Gell-Mann, P.~Ramond and R.~Slansky, in
Supergravity, eds.\ P.~van~Nieuwenhuizen and D.Z.~Freedman
(North Holland, Amsterdam, 1979); R.~N.~Mohapatra and G.~Senjanovic, 
 Phys.\ Rev.\ Lett.\  {\bf 44} (1980) 912;
J.~Schechter and J.~W.~F.~Valle,
Phys.\ Rev.\ D {\bf 22} (1980) 2227; Phys.\ Rev.\ D {\bf 25} (1982) 774.

\bibitem{Pilaftsis:2003gt}
  A.~Pilaftsis and T.~E.~J.~Underwood,
  Nucl.\ Phys.\ B {\bf 692} (2004) 303
  [hep-ph/0309342].

\bibitem{Kuzmin:1985mm}
  V.~A.~Kuzmin, V.~A.~Rubakov and M.~E.~Shaposhnikov,
  Phys.\ Lett.\ B {\bf 155} (1985) 36.






\bibitem{Kersten:2007vk}
  J.~Kersten and A.~Y.~.Smirnov,
  Phys.\ Rev.\ D {\bf 76} (2007) 073005
  [arXiv:0705.3221 [hep-ph]].

\bibitem{Ibarra:2011xn}
  A.~Ibarra, E.~Molinaro and S.~T.~Petcov,
  Phys.\ Rev.\ D {\bf 84} (2011) 013005
  [arXiv:1103.6217 [hep-ph]].

\bibitem{Dinh:2012bp}
  D.~N.~Dinh, A.~Ibarra, E.~Molinaro and S.~T.~Petcov,
  JHEP {\bf 1208} (2012) 125
  [arXiv:1205.4671 [hep-ph]].

\end{thebibliography}
\end{document}